# Black and White Anatase, Rutile and Mixed Forms: Band-Edges and Photocatalytic Activity

Xuemei Zhou, Ewa Wierzbicka, Ning Liu, Patrik Schmuki*

Department of Materials Science, WW4, LKO, University of Erlangen-Nuremberg, Martensstrasse 7, 91058, Erlangen, Germany;

*Corresponding author. Tel.: +49 91318517575, Fax: +49 9131 852 7582 Email: schmuki@ww.uni-erlangen.de.

Link to the published article:

https://pubs.rsc.org/en/content/articlehtml/2019/cc/c8cc07665k

**Here we investigate the band–level energetics of "black" hydrogenated titania in different polymorphs using in-situ photoelectrochemical measurements and XPS valence band measurements. We find that the conduction band of black rutile is higher in energy than in black anatase by 0.4 eV. For photocatalytic hydrogen generation, in a polymorph hetero-junction such as in black Degussa P25, thus black rutile can act as a photosensitizer while black anatase provides charge- mediation catalysis onto $H_2O$ to generate $H_2$. By optimizing the thermal reduction conditions of black anatase/rutile junctions the $H_2$ production can be significantly increased.**

Since 1972 when Fujishima and Honda discovered the photocatalytic water splitting ability of titania,[1] $TiO_2$ has become a benchmark photocatalyst. One of the crucial features of titania that make it a photocatalyst for $H_2$ evolution are the relative energetic positions of the conduction and valence band relative to the redox potential ($E_{redox}$) of water.[2] However, the band edge positions vary for different polymorphs of titania and thus the thermodynamic driving force for water reduction is different. For anatase and rutile, as well as for junctions formed by two phase photocatalysts (such as Degussa P25) significantly different band-edge positions and alignments have been reported, often depending on the techniques used to characterize the material.[3-9] Typically electrochemical approaches and photoemission techniques, such as XPS/UPS measurements are used to investigate this question. Even more complex becomes the situation if other modifications of titania are regarded.

Recently, a large body of research work has been dedicated to anatase converted by hydrogenation to so-called "black titania"[10-12] – this modification also implies alterations in the energy positions of the band-edges. "Black anatase" has attracted wide scientific attention due to *i)* an extension of the absorption spectra of titania to the visible range,[13-16] and *ii)* applications as intrinsically activated photocatalysts for open circuit hydrogen evolution without the use of noble-metal co-catalysts – the latter form of titania is often termed "grey" anatase due to milder reduction condition used to obtain this modification.[17-18] Liu et al. [19] reported that only anatase, and not rutile, can be activated to provide an intrinsic co-catalytic center that facilitates $H_2$ generation.

"Black" and "grey" $TiO_2$ show different properties from white titania, e.g. an improved conductivity, formation of amorphous shells around crystalline $TiO_2$ particle, and band edge tailing (reason for visible light absorption), among others.[10,18]

However, the energetic of the band positions of black titania polymorphs have hardly been investigated, even less when the two black polymorphs are present as a hetero-junction, such as in black mixed phase material (P25). Characterization of the co-catalytically active black anatase by PL showed the formation of shallow electronic states below the conduction band.[19] These states were found to be a mediator for the transfer of electrons from the conduction band of anatase to the electrolyte (i.e. they act as intrinsic co-catalytic centers for $H_2$ evolution similar to noble metals).[20] However, up to now, there are no reports on the investigation of the band-edge positions of hydrogenated rutile or mixed phase titania, although, mixed phase hydrogenated titania were reported to provide a higher photocatalytic activity than grey anatase.

In the present work, we use a combination of in-situ photoelectrochemical measurements and XPS measurements on defined anatase and rutile electrodes (in their black and white form) to extract information on changes in their electronic structure (band gap, energetic positions of conduction and valence band) due to reduction. Experiments are performed using different forms of titania, including nanoparticle (NP) electrodes and defined single crystal (SC) substrates. We furthermore investigate anatase/rutile mixed particles such as variations of black P25, obtained under different high temperature hydrogen annealing conditions, and examine the photocatalytic $H_2$ generation for different mixed phase anatase to rutile ratios. Within the manuscript, the samples are denoted as X-NP/SC-$H_2$-T, where X stands for anatase, rutile or P25, NP for nanoparticle, and SC for single crystal, T represents the treatment temperature, and the hydrogenated materials are termed "black".

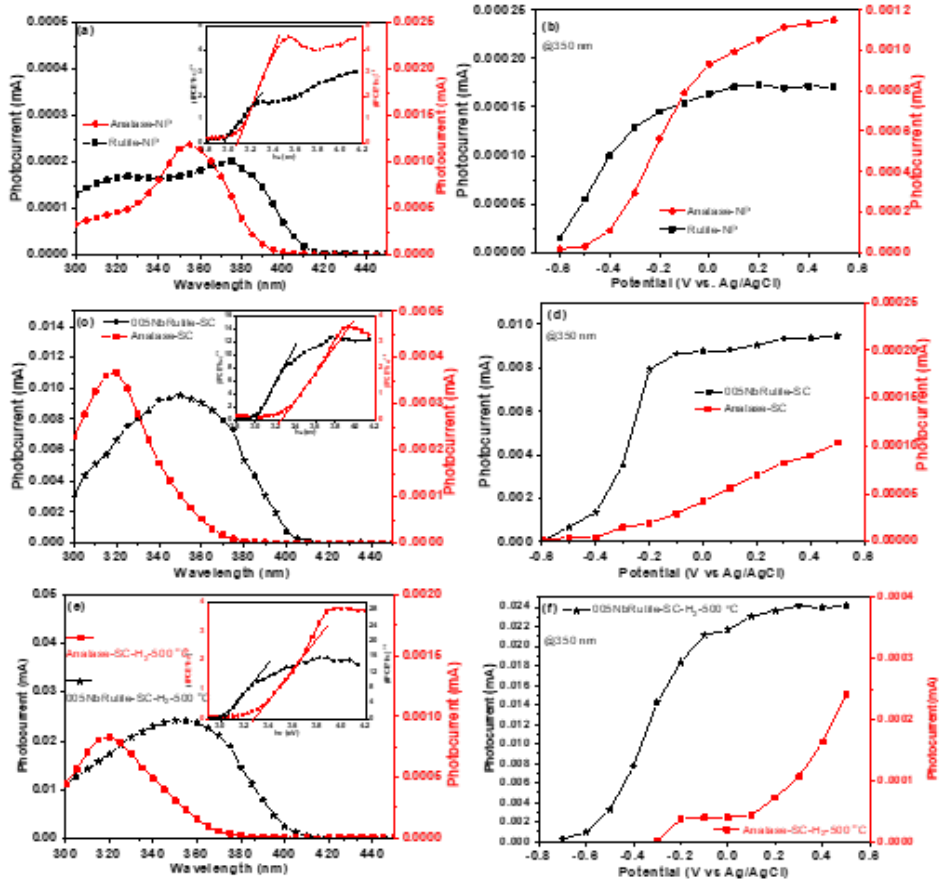

*Figure 1.* Photocurrent measurements on anatase and rutile electrodes made from (a) powders, (c) single crystals, and (e) hydrogenated anatase and rutile phases. Inset of (a), (c) and (e) are band gap evaluations from photocurrent spectroscopy for (a) powders, (c) single crystals and (e) hydrogenated phases. Photocurrents in function of applied potential for (b) anatase and rutile electrodes made from powders, (d) single crystals, and (f) hydrogenated anatase and rutile phases.

**Fig. 1** shows photocurrent spectra and photoelectrochemical I-V curves of anatase and rutile powder, as well as for single crystal substrate electrodes. The SEM images in the **supporting information (Fig. S1)** show a uniformly coated layer of anatase powders on the FTO substrates. Both anatase and rutile electrodes show a photocurrent response in the UV range. In our experiments, the rutile powder electrodes show a lower photocurrent than anatase powder electrodes (**Fig. 1a**) – this we ascribe to the intrinsically lower electrical conductivity of rutile [21]. The band gap for the anatase powder electrode is evaluated to be 3.08 eV while for the rutile a value of 2.95 eV is determined (**Fig. 1a inset**). Photoelectrochemical I-V curves were scanned from 0.5 V to -0.6 V at a fixed wavelength of 350 nm. The photocurrent onset potential, that is the photoelectrochemical flatband potential ($U_{fb}$), is approximately -0.5 V (vs. Ag/AgCl) for anatase NPs and approx. -0.6 V (vs. Ag/AgCl) for rutile NPs (**Fig. 1b**). From these measurements one may conclude that for the powder electrodes, the CB edge of rutile lies 0.1 eV higher in energy than for anatase.

In a second approach, we used anatase and rutile single crystal (SC) substrates (both of a [001] orientation). In the case of rutile, doped single crystals of a high conductivity are commercially available (0.05 wt.% Nb-doped) which exclude conductivity issues in photoelectrochemical experiments. Photocurrent spectra for the doped rutile single crystal (**Fig. 1c**) allow to extract a band gap ($E_g$) (**Fig. 1c inset**) of 3.01 eV. The anatase SC was a naturally grown crystal with a low intrinsic conductivity – however still small photocurrents could be measured. For this photoelectrode we obtained a band gap of 3.28 eV – thus these $E_g$ data are well in line with literature for rutile and anatase.[22]

From the photoelectrochemical I-V curves (**Fig. 1f**), the onset potential for the doped rutile SC is -0.6 V (vs. Ag/AgCl) while for the anatase SC an onset was measured at -0.4 V (vs. Ag/AgCl). These values of $E_g$ and of $U_{fb}$ are very similar to the results obtained for powder electrodes (**Fig. 1d**), i.e. confirm also for the single crystals that the CBM of rutile SC is positioned 0.2 eV higher in energy than anatase.

The same set of experiments was performed after hydrogenating the single crystal substrates – results are shown in **Fig. 1e**. In photocurrent measurements, after hydrogenation of the doped rutile SC at 500 °C, the photocurrent increased (**Fig. 1e**), while the band gap (**Fig. 1e inset**) did not change significantly ($E_g$ = 3.00 eV). However, the photocurrent onset potential (**Fig. 1f**) shifted to -0.7 V (vs. Ag/AgCl), which is 0.1 eV higher in energy than the untreated doped rutile SC.

For the hydrogenated anatase SC (Hy-AnSC), the band gap evaluation yields 3.25 eV. However, the photocurrent onset for the hydrogenated single crystal is located at -0.3 V (vs. Ag/AgCl), which is 0.1 eV lower in energy than the value for the untreated anatase SC (**Fig. 1f**). Please note that after the hydrogenation treatment, surface hydroxyl groups can be observed on the anatase surface with a clearly higher concentration than on the rutile surface (see XPS spectra in **Fig. S2**). In this context it is interesting to note that earlier investigations using PL showed for hydrogenated anatase the presence of sub-band gap states approx. at 0.2 eV below the conduction band of anatase.[20,23]

Based on above analysis (band gap values and conduction band position), we propose a band alignment of hydrogenated black anatase and rutile as shown in **Fig. 2a**, where the conduction band of black rutile lies 0.4 eV higher in energy than that of black anatase.

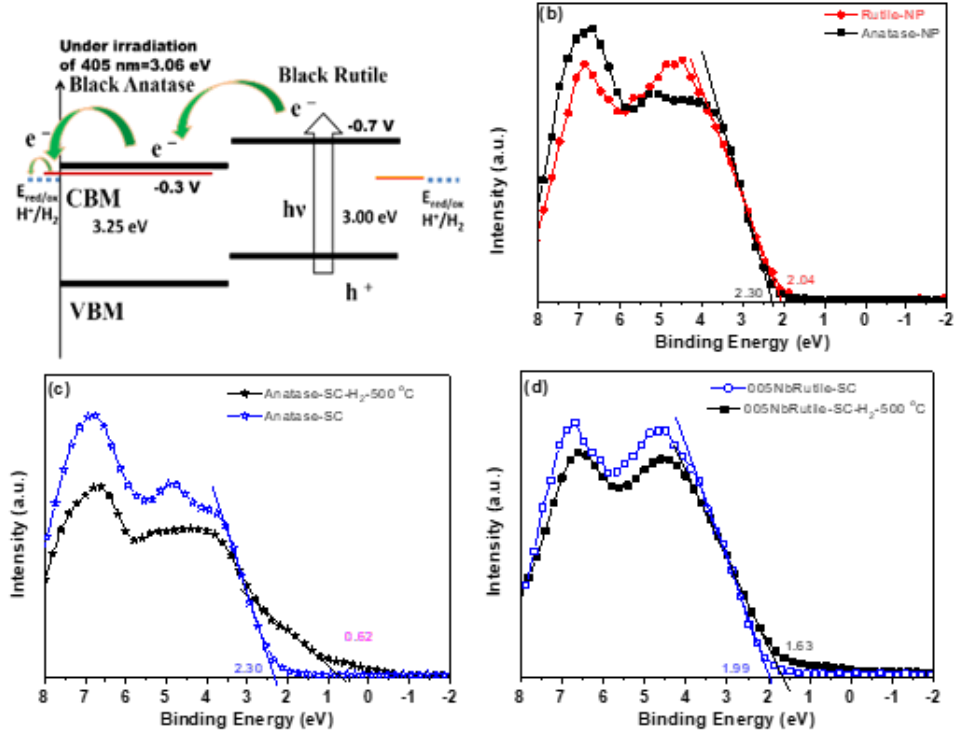

*Figure 2*. *(a) Schematic drawing of the energy diagrams between hydrogenated anatase and rutile under illumination of light at λ= 405 nm. (b) XPS valence band measurements for anatase and rutile electrodes made from powders. (c) XPS valence band measurements for anatase and hydrogenated anatase electrodes made from single crystals. (d) XPS valence band measurements for rutile and hydrogenated rutile electrodes made from single crystals.*

To further characterize the electronic structure for white and hydrogenated black titania, we performed XPS valence band measurements for the powder samples. For rutile samples we find that the VBM for rutile lies at a higher energy than anatase, e.g. 2.30 eV for anatase NPs and 2.04 eV for rutile NPs. (**Fig. 2b**) That is, the VBM of rutile is around 0.2 - 0.3 eV higher in energy than that of anatase. This is in line with theoretical data from Liang et al.[24] that for the most stable heterojunction of (white) anatase and rutile calculated a valence band maximum (VBM) of rutile that lies 0.52 eV above that of anatase. [25]

For the black samples: the hydrogenated anatase single crystal shows a tail of VBM around 0.62 eV, which is 1.70 eV higher in energy than VBM of anatase single crystal (2.32 eV) (**Fig. 2c**) – this is well in line with literature data for the VBM of anatase after hydrogenation. [10] But also for the black rutile SC, we find a shift of the VBM spectrum (located at 1.64 eV), which is 0.33 eV higher in energy than the untreated rutile single crystal (**Fig. 2d**). However, these states are not apparent in the photocurrent spectra (no lowering of the band gap is observed). This indicates that photogenerated charge carriers trapped on such states show a very low mobility.

In Degussa titania P25, both anatase and rutile particles exist as agglomerates with 75% anatase and 25% rutile.[26] For P25 considerable discussion exists, if the two phases are electronically coupled, i.e., if they indeed form an electronic hetero-junction. Such heterojunctions have often been identified to be the origin of a frequently reported improved photocatalytic activity for P25. Here we compared black anatase, black rutile and black P25 for their intrinsic photocatalytic $H_2$ generation that is without the use of any external co-catalyst. We use different illumination conditions: UV light, AM1.5 solar light (100 mW/cm$^2$) and a 405 nm laser (3.06 eV), that can only excite the rutile band gap of 3.0 eV but not the anatase band gap of 3.2 eV.

First it should be noted that the $H_2$ treatment of P25 causes a shift in the anatase to rutile ratio. XRD results in **Fig. 3a** show an increase of the rutile content in black P25 for higher hydrogenation temperatures. For UV illumination (**Figure 3b**), where both anatase and rutile can be excited, it is evident that all non-hydrogenated samples show only a minimal amount of hydrogen generation. White anatase and P25 show a slightly higher $H_2$ evolution activity than white rutile. After hydrogenation, all materials become more active but only anatase and P25 become significantly enhanced regarding $H_2$ evolution. Depending on the annealing conditions, P25 can obviously reach an activity that is considerably higher than the best anatase results. Under UV illumination, an optimum condition for black P25 material is for annealing at 450 °C, where anatase is the main phase.

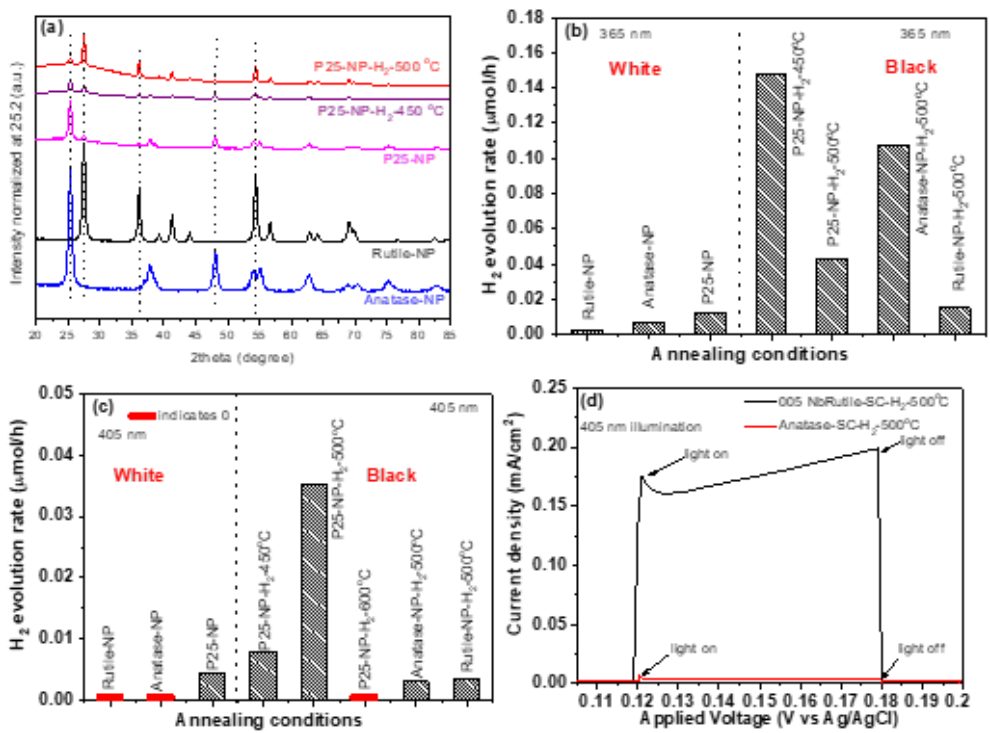

*Figure 3.* (a) XRD for anatase, rutile, P25 and hydrogenated P25. (b) Photocatalytic $H_2$ evolution of these materials under illumination of light at λ= 365 nm without the use of noble-metal co-catalysts. (c) Photocatalytic $H_2$ evolution of these materials under illumination of light at λ= 405 nm without the use of noble-metal co-catalysts. (d) Photoelectrochemical measurements for (a) hydrogenated rutile single crystal and (b) hydrogenated anatase single crystal electrodes under illumination of light source of λ=405 nm.

The situation becomes different if 405 nm light is used for illumination (as this wavelength enables the excitation of rutile only). Here white anatase as well as white rutile are inactive (**Fig. 3c**). However, white P25 is able to generate a small amount of $H_2$, which clearly indicates that the mixed-phase has a beneficial effect (most likely due to a better charge carrier separation[27]). Most interesting however is that black P25 (treated at 500 °C) generates a much higher amount of hydrogen than the P25 treated at 450 °C and white P25. This indicates that with a higher content of rutile in the junction, a higher conversion efficiency of incident light to H$_2$ can be obtained, suggesting that rutile in this case acts as an antenna to absorb 405 nm light. Then the photoexcited electrons in the conduction band of rutile are transferred to anatase and finally react via a Ti$^{3+}$ surface state (**Fig. 2a**) with the electrolyte.[28]

To support that only rutile can be excited with an illumination of 405 nm light, we performed additional photoelectrochemical measurements for hydrogenated single crystal substrates under 405 nm (**Fig. 3d**). The photocurrent for hydrogenated Nb-doped rutile single crystal is very high >0.15 mA/cm$^2$, compared with hydrogenated anatase single crystal <0.01 mA/cm$^2$). For the plain anatase crystal no signal is detectable. The difference between hydrogenated anatase and plain anatase reflects the presence of the sub-gap states in hydrogenated anatase that can act as electron transfer mediator to the liquid environment, as described in the literature. [18,29]

**Conclusions**

In summary, in this work we investigated the electronic structure of white anatase and rutile, and compared it to black anatase and rutile. For this we used photoelectrochemical techniques and VB measurements using XPS. For white samples our data are in line with various reports,[7,8,30,31] i.e. the CBM of rutile lies slightly higher in energy than CBM of anatase and so do the valence band positions. After hydrogenation, the band edge positions change by -0.1 eV for anatase and 0.1 eV for rutile. In anatase, sub-band gap states are present as illustrated in red in **Figure 2a**.

For P25 clearly a beneficial junction is formed − the photocatalytic experiments show that the photogenerated electrons can transfer from the CBM of rutile to the CBM of anatase. In other words: for P25, rutile plays the role of a sensitizer (absorber) for light harvesting, anatase provides the charge transfer center needed to generate $H_2$. The present work not only provides data on the energetics in "black" $TiO_2$ polymorphs but also elucidates the different roles of titania polymorphs in their "black" mixed phase forms, and explains their roles in photocatalytic $H_2$ generation in absence of any external (noble metal) co-catalyst.

**Conflicts of interest**

There are no conflicts to declare.

**Acknowledgements**

The authors would like to acknowledge ERC, DFG, the Erlangen DFG cluster of excellence (EAM) for their financial support.